\documentclass[twocolumn,pre,showpacs,preprintnumbers,amsmath,amssymb]{revtex4}

\usepackage{graphicx}
\usepackage{dcolumn}
\usepackage{bm}
\usepackage{epsf}
\usepackage{mathrsfs}

\begin{document}
\title{Realization of Discrete Quantum Billiard in 2D Optical Lattices}
\author{Dmitry O. Krimer${}^{1,2}$}\email{dmitry.krimer@gmail.com}
\author{Ramaz Khomeriki${}^{2,3}$}\email{khomeriki@hotmail.com}
\affiliation{ ${\ }^1$Theoretische Physik, Universit\"at T\"ubingen, Auf der
Morgenstelle 14, 72076 T\"ubingen, Germany \\
${\ }^2$Max-Planck Institute for the Physics of
Complex Systems,
N\"othnitzer Str. 38, 01187 Dresden, Germany \\
${\ }^3$Physics
Department, Tbilisi State University, Chavchavadze 3, 0128
Tbilisi, Georgia}

\begin{abstract}

We propose the method for optical visualization of Bose-Hubbard
model with two interacting bosons in the form of two-dimensional (2D)
optical lattices consisting of optical waveguides, where the
waveguides at the diagonal are characterized by different
refractive index than others elsewhere, modeling the boson-boson
interaction. We study the light intensity distribution function
averaged over direction of propagation for both ordered
and disordered cases, exploring sensitivity of the averaged
picture with respect to the beam injection position. For our
finite systems the resulting patterns reminiscent the ones set in
billiards and therefore we introduce a definition of discrete
quantum billiard discussing the possible relevance to its well
established continuous counterpart.
\end{abstract}
 \maketitle

A very rich variety of wave phenomena that have originally been
discovered in the context of atomic and solid state physics
attracted recently much attention due to their deep analogy to
optical systems.  A first prominent example is the Anderson
localization, the phenomenon which was originally discovered as
the localization of electronic wavefunction in disordered crystals
\cite{PWA58} and later understood as a fundamental concept being
universal phenomenon of wave physics. Related recent experiments
were performed on light propagation in spatially random nonlinear
optical media \cite{Exp, Exp2} and on Bose-Einstein condensate
expansions in random optical potentials \cite{BECEXP}. A second
example is the well known solid state problem of an electron in a
periodic potential with an additional electric field, which lead
to investigations of Bloch oscillations and Landau-Zener tunneling
in various physical systems such as ultracold atoms in optical
lattices \cite{BEC_lin,BEC_nonlin,cold_exp} and optical waves in
photonic lattices \cite{OL_lin,OL_nonlin}. A recent progress in
the experiments stimulated a new turn in theoretical studies
dealing with the evolution of a wave packet in nonlinear
disordered chains \cite{Anders_non}, in a nonlinear Stark ladder
\cite{Stark_nonlin} and the effect of Anderson localization of
light near boundaries of disordered photonic lattices
\cite{JKDB11} which are just a few recent examples. A third very
interesting example is a classical analog of beam dynamics in one-dimensional (1D)
photonic lattices to quantum coherent and displaced Fock states
\cite{PCSC10} and a classical realization of the two-site
Bose-Hubbard model (applicable to the physics of strongly
interacting many-body systems), based on light transport in
engineered optical waveguide lattices \cite{L11}.
%
\begin{figure}[b]
\begin{center}
\rotatebox{0}{\includegraphics[width=4.5cm]{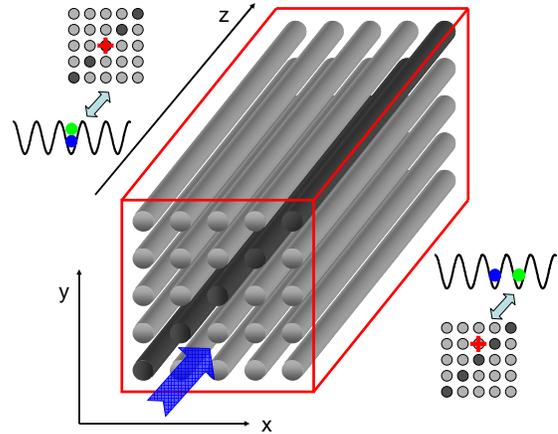}}
\end{center}
\caption{Geometry of
setup: A beam enters into 2D optical lattice, and
propagates along the $z$ axis. The refractivity index is invariant
along the $z$ axis and either periodic or disordered in transverse directions. The corresponding mapping to the dynamics of two interacting distinguishable bosons in a chain
is also done (see the text for details). The interaction between
bosons is introduced by taking the refractive index for the
diagonal waveguides different from the rest. The injection of a
beam to the diagonal waveguide mimics lunching initially both
bosons at the same site (upper inset), while injecting the beam
into off-diagonal waveguide corresponds to the two bosons located
initially on different sites (lower inset).}
\label{fig_arrays}\end{figure}
%

In this Letter we study a classical analog of beam propagating in
2D photonic lattices to quantum coherent dynamics of two particles
in one-dimensional chain using the Bose-Hubbard model. We consider
different situations ranging from the simple ordered case without
interaction to the disordered case with interaction in our finite
systems. Sometimes the resulting patterns look pretty similar as
the ones for the classical and/or quantum billiards which are known
to exhibit regular and chaotic behaviors (see e.g. Ref.
\cite{stockman}). We would like to emphasize particularly the
growing interest to the two-particle problem in the context of
quantum correlations between two noninteracting particles evolving
simultaneously in a disordered medium \cite{LBCS10} and quantum
walks of correlated photons which provides a route to universal
quantum computation \cite{quant_inform}. Thus, the obtained
results might be applicable to both classical and quantum systems.

Let us introduce a standard Bose-Hubbard Hamiltonian describing
two distinguishable bosons (or two fermions with opposite spins)
in a chain with $N$ sites
\begin{equation}
{\cal \hat H}= \sum\limits_{j=1}^N\left[\left(\hat a_{j+1}^+\hat
a_{j}+\hat b_{j+1}^+\hat b_{j}+h.c.\right)+U\hat a_{j}^+\hat
a_{j}\hat b_j^+\hat b_j\right] \label{eq1}
\end{equation}
where $\hat b_{j}^+$ ($\hat a_{j}^+$) and $\hat b_{j}$ ($\hat
a_{j}$) are bosons creation and annihilation operators on a lattice
site $j$ and $U$ is the onsite interaction strength. Starting from
the time dependent Schr\"odinger equation $i\partial_t
|\Psi(t)\rangle={\cal \hat H} |\Psi(t)\rangle$ we expand
$|\Psi(t)\rangle$ in terms of the $N^2$ orthonormal eigenstates of
a number operator, $|m,n\rangle\equiv\hat b_{m}^+\hat
a_{n}^+|0\rangle$, as
$|\Psi(t)\rangle=\sum\limits_{m,n=1}^{N}c_{mn}(t)|m,n\rangle$, where
the amplitudes $c_{mn}(t)$ satisfy to the following set of
equations
%
\begin{figure}[t]
\hspace*{-1.2cm}
\rotatebox{0}{\includegraphics[width=11.cm]{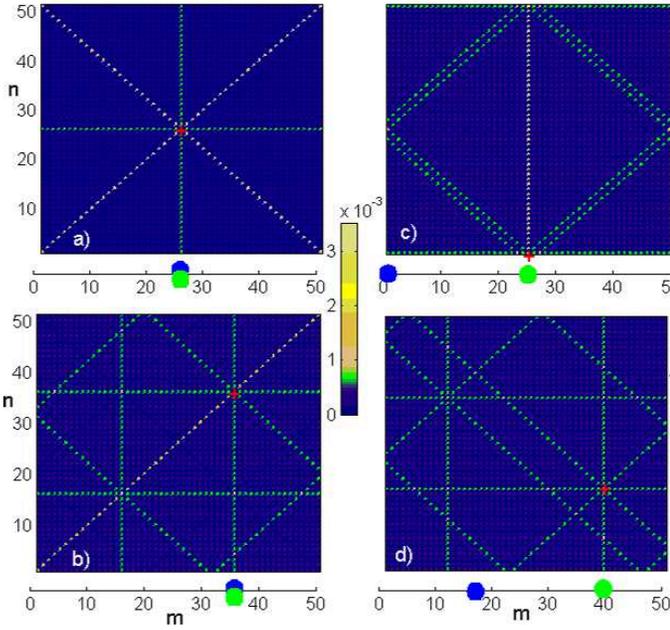}}
\caption{(Color online) Characteristic pictures for discrete quantum billiard realization
for different injection points depicted by a red cross in the
absence of disorder and interaction ($W=0$ and $U=0$). In the main
graphs the averaged PDFs [see Eq.~\eqref{eq3}]  are displayed and
accompanied by the lower insets which show the numbers of initially injected waveguides
(or number of sites at which the particles are initially located).} \label{hubbardsurf1}\end{figure}
%
%
\begin{eqnarray} \label{eq2}
i \dot c_{mn}=U\delta_{mn}c_{mn}+\sum\limits_{m',n'=1}^{N}R_{mn}^{m'n'}c_{m'n'},
\quad R_{mn}^{m'n'}= ~~~~
\\
\delta_{m'm+1}\delta_{n'n}+\delta_{m',m-1}\delta_{n'n}+
\delta_{m'm}\delta_{n'n+1}+\delta_{m'm}\delta_{n'n+1} \nonumber.
\end{eqnarray}
Note that Eq.~(\ref{eq2}) is invariant under permutation of
$m$ and $n$, and therefore it is natural to represent
$c_{mn}$ as a sum of symmetric $c_{mn}^S=(c_{mn}+c_{nm})/\sqrt{2}$
and antisymmetric $c_{mn}^A=(c_{mn}-c_{nm})/\sqrt{2}$ functions.
In such a basis the matrix $R_{mn}^{m'n'}$ is decomposed into two
irreducible parts, one of which corresponds to the Bose-Hubbard
model with two indistinguishable bosons and the other describes
the physics of two indistinguishable spinless fermions. For the
symmetric initial conditions, $c_{mn}(0)=c_{nm}(0)$, the dynamics
is reduced to the former case (two indistinguishable bosons
on sites $m$ and $n$), whereas the latter case is
realized for the antisymmetric initial conditions,
$c_{mn}(0)=-c_{nm}(0)$.
%
\begin{figure}[t]
\hspace*{-1.2cm}
\rotatebox{0}{\includegraphics[width=11.cm]{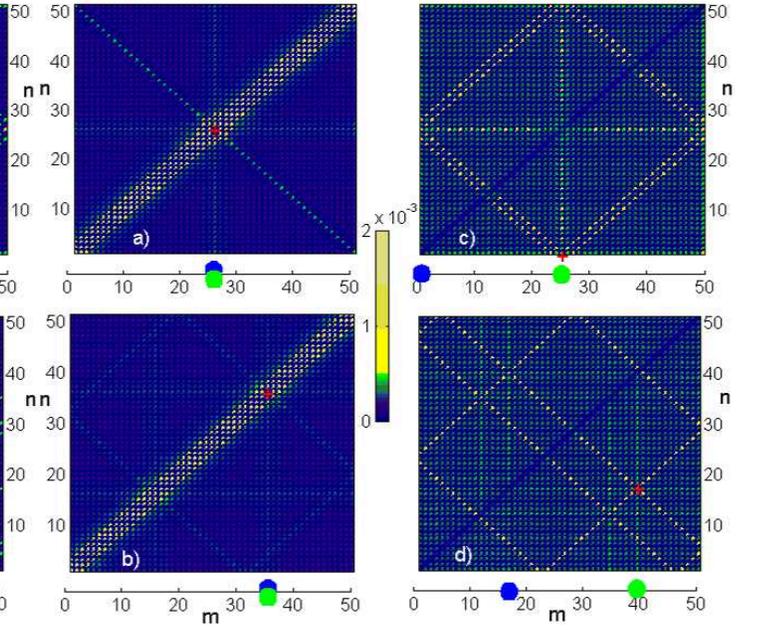}} 
\caption{(Color online)
All parameters and quantities are the same as in
Fig.~\ref{hubbardsurf1}, except the interaction constant $U=1$.}
\label{hubbardsurf2}\end{figure}

Remarkably, Eq. \eqref{eq2} is the same as one used for the
description of light propagation through 2D optical lattices (see
Fig.~\ref{fig_arrays}) within the tight-binding approximation,
where longitudinal dimension $z$ plays a role of time. This
approximation is valid when a lattice is constructed such that
tunneling into nearest neighboring waveguides is allowed only and
there is a difference between the refractive indices of the
diagonal $n_d$ and off-diagonal $n_0$ waveguides which models the
interaction (with the interaction strength $U\sim n_0-n_d$). Thus,
injecting light beam at the waveguide with a position $x=m$, $y=n$
(asymmetric initial conditions) corresponds to  the dynamics of
two distinguisable interacting bosons in a chain,
placed initially on sites $m$ and  $n$. One can also think about
the Bose-Einstein condensate embedded into 2D optical lattice and
then Eq.~\eqref{eq2} describes the evolution of some initial
matter wave packet through the lattice.

In this Letter we consider the system with hard boundaries having $c_{mn}=0$ outside a square
and monitor the time averaged wave function
\begin{equation} \label{eq3}
{P}_{mn}\equiv \lim_{T\rightarrow \infty} \frac{1}{T}\int_0^T |c_{mn}(t)|^2dt,
\end{equation}
referring to $P_{mn}$ as to the averaged two-particle probability distribution function (PDF). To calculate PDFs
we, at first, solve the eigenvalue problem ${\cal \hat H}|q\rangle=\lambda_q |q\rangle$ and then
expand $c_{mn}(t)$ with respect to the eigenvectors as
\begin{equation} \label{eq4}
c_{mn}(t)=\sum_{q=1}^{N^2}\phi_q{\cal L}_{mn}^{(q)} e^{-i\lambda_q t},
\end{equation}
where ${\cal L}_{mn}^{(q)}\equiv\langle q|m,n\rangle$ is the eigenvector which belongs to the eigenvalue
$\lambda_q$ and $\phi_q\equiv\sum_{m,n=1}^N c_{mn}(0){\cal
L}_{mn}^{(q)}$ is its initial amplitude. Next, the averaged PDF is calculated by
the following formula:
\begin{equation} \label{eq5}
P_{mn}=\sum_{q}\left|\varphi_{q}\right|^2
{\cal L}_{mn}^{(q)2}+\sum_i\left|\sum_{q_i^r}\varphi_{q_i^r}
{\cal L}_{mn}^{(q_i^r)}\right| ^2,
\end{equation}
where the first sum runs over all nondegenerate eigenvalues,
whereas the second sum corresponds to the summation with respect to
$r$-fold degenerate eigenvalues $\lambda_{q_i}$.
\begin{figure}
\hspace*{-1.2cm}
\rotatebox{0}{\includegraphics[width=11.cm]{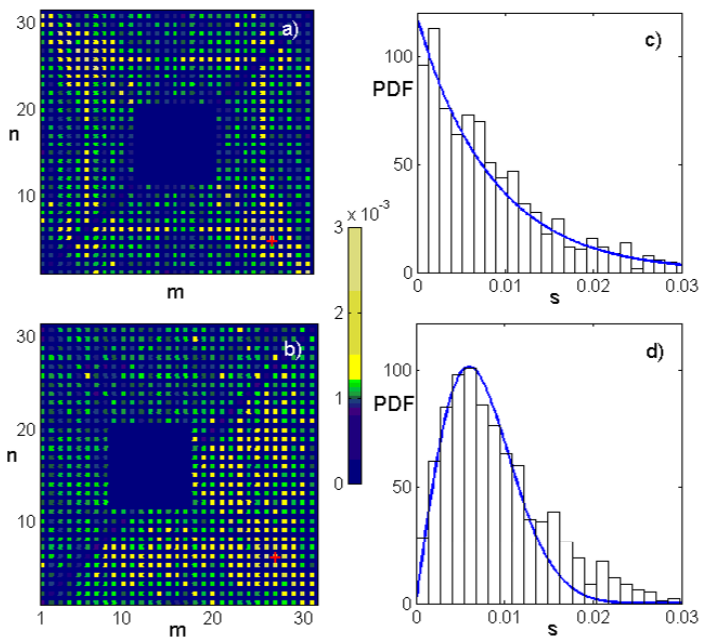}} 
\caption{(Color online)
Characteristic pictures for discrete quantum billiard realization
with a rigid square placed inside a system for the interaction
constant $U=1$ and the same injection point depicted by a red
cross. (a),(b): symmetric and asymmetric situations, respectively,
with the corresponding probability density functions of eigenvalue
spacings $s$ shown in (c) and (d). (c):  the Poisson distribution
\eqref{poisson}, with the average spacing $d=0.0085$. (d):  the
Wigner-Dyson distribution \eqref{wigner}, with the average spacing
$d=0.0075$.} \label{hubbardsurf3}\end{figure}
%

Intuitively it seems that the light injected into one of the
waveguides should spread over a whole lattice, however the real
situation is completely opposite due to the interference from the hard
boundaries. Let us start form the simplest noninteracting case,
$U=0$ (for the optical counterpart shown in Fig.~\ref{fig_arrays}
waveguides must all be identical). As is seen from
Fig.~\ref{hubbardsurf1} a well defined pattern for $P_{mn}$
corresponds to each initial injection point. {\it Thus, the system
keeps the information about its initial state and from the
averaged picture one can recover an initial signal.} It should be
noted, that these patterns might be strongly modified when the
interaction is switched on, $U \neq 0$ (see
Fig.~\ref{hubbardsurf2}). Remarkably, the pattern's structure
reminiscent the one sets in billiards, therefore, we introduce the
notion of discrete quantum billiard and seek for the analogies
with usual continuous counterparts. The first step towards this
direction is to explore the possibility of quantum chaos
realization in such systems. We consider two possibilities to
observe the transition towards quantum chaos. The first one is a
symmetry breaking by placing a square with rigid boundaries inside
the system as is shown in Fig.~\ref{hubbardsurf3}. We monitor then
the statistical properties of the eigenvalue spacings
$s=|\lambda_{q+1}-\lambda_{q}|$ for different locations of the
square, keeping injection point and interaction constant the same.
It is seen that in the symmetric case the Poisson distribution
\begin{equation}\label{poisson}
P(s)=1/d\cdot e^{-s/d}
\end{equation}
is realized, while for the asymmetric case the Wigner-Dyson distribution is observed
\begin{equation}\label{wigner}
P(s)=\pi s/(2 d^2)\cdot e^{-\pi s^2/(4d^2)}.
\end{equation}
Thus the onset of quantum chaos could be visualized via the classical optical system of
coupled waveguides.

The second mechanism of quantum chaos realization is an introduction of the disorder via
adding the following terms to the Bose-Hubbard Hamiltonian
\eqref{eq1}
\begin{eqnarray}\label{eq6}
{\cal \hat H}_d= \sum_{j=1}^N\left[\epsilon_{j}^a\hat a_{j}^+\hat
a_{j}+\epsilon_j^b\hat b_{j}^+\hat b_{j}\right],
\end{eqnarray}
where for the sake of simplicity we take symmetric disorder
$\epsilon^a_j=\epsilon^b_j\equiv\epsilon_j$ ($\epsilon_j$ are
random numbers from the interval $[-W/2,W/2]$, $W$ being the
disorder strength). In the optical context presented in
Fig.~\ref{fig_arrays} it implies the usage of symmetric (under permutation of $m$
and $n$) disordered lattice described by the following modified evolution equations:
\begin{equation} \label{eq7}
i \dot c_{mn}=\left(W_{mn}+U\delta_{mn}\right)c_{mn}+
\sum\limits_{m',n'=1}^{N}R_{mn}^{m'n'}c_{m'n'},
\end{equation}
where the matrix $R_{mn}^{m'n'}$ is the same as in Eq.~\eqref{eq2}
and $W_{mn}$=$\epsilon_m+\epsilon_n$ are correlated disorder
parameters. An uncorrelated disorder can also be taken with
$W_{mn}$ being a sum of two independent random numbers for each $m$ and $n$. In the optical
context such a situation might be realized by taking either
correlated or uncorrelated random refractive index distribution in
a whole lattice. Note, that only the former case coincides
with the dynamics of two bosons in a disordered chain with a
random potential.

%
%
\begin{figure}
\hspace*{-1.2cm}
\rotatebox{0}{\includegraphics[width=11.cm]{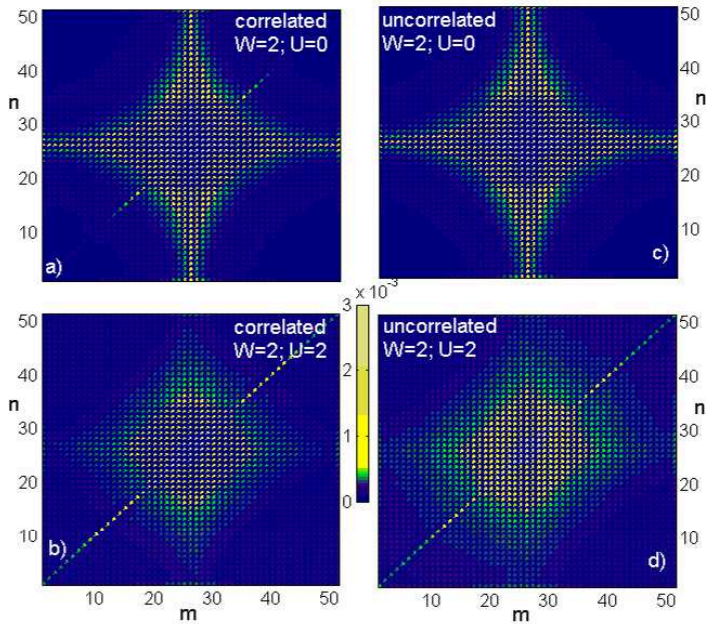}} 
\caption{(Color online)
Averaged PDFs both in time and over many disorder realizations
(all parameters are shown on the figures). Injection point in all
graphs is taken at the middle of 2D optical lattice.}
\label{fig_Wign}\vspace*{.5cm}\end{figure}
%
\begin{figure}[t]
\hspace*{-1.2cm}
\rotatebox{0}{\includegraphics[width=5.cm]{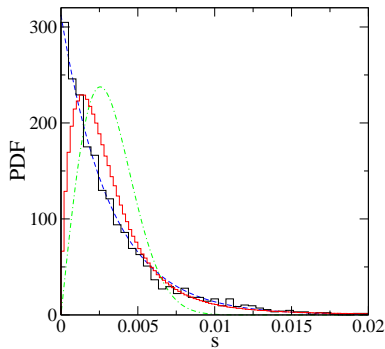}} 
 \caption{ (Color online)
Probability density functions of eigenvalue spacings $s$ for three
different cases and a chain with $N=51$. Black curve: W=0, U=2
(disorder strength is zero only). Red curve (dark gray): W=2, U=2.
The case with uncorrelated disorder is considered. Green
dash-dotted curve: the Wigner-Dyson distribution \eqref{wigner},
with the average spacing $d=0.0032$. Blue dashed curve: the
Poisson distribution \eqref{poisson}, with the average spacing
$d=0.0032$.} \label{fig_statitst}\end{figure}
%

We consider two particles which are initially launched on the same
site at the middle of a chain, $m=n=N/2$ (the optical counterpart corresponds to the beam
injection at the central waveguide in 2D optical lattices). 
The typical structures for $P_{mn}$, averaged out with respect to many disorder realizations,
are shown in Fig.~\ref{fig_Wign}. As is seen, the averaged
PDFs demonstrate well pronounced patterns, which look differently as compared to
the case with a single disorder realization, when PDF has many
spots at different locations. For the noninteracting case PDF has
an anisotropic structure with two distinct directions, $m=N/2$ and
$n=N/2$, along which the particle motion mostly develops in
average. For the case with correlated disorder, a slight
contribution of two particle states is also visible. For $U=2$,
the interaction is already strong enough, such that the
contribution of states corresponding to the breather band becomes
essential and the two particles mostly prefer to form a composite
state and travel together. Interestingly, the level of system's chaoticity might easily be governed
by setting different values of disorder and interaction
strengths. To demonstrate this we study the statistical properties
of the eigenvalue spacings $s=|\lambda_{q+1}-\lambda_{q}|$ for
three different cases as is shown in Fig.~\ref{fig_statitst} (the
events with $s=0$ due to degeneracy are not counted). It is seen
that for nonzero disorder and  interaction strengths, probability
density function of the spacings $s$ has a tendency to go to the
Wigner-Dyson distribution and, as a consequence, a system becomes
more chaotic.

Concluding, in this paper we have discussed various interpretation
of the optical beam propagation through 2D optical crystal ranging
from interacting cold atom dynamics and two particle Anderson
localization to the quantum billiard problems connected with
transition to quantum chaoticity.

{\it Acknowledgements:} The authors are indebted to I. Babushkin, S. Denisov and N.
Li for usefull discussions regarding the Billiard issues. R.Kh. is
supported by RNSF (Grant No. 09/04) and STCU (Grant No. 5053).

\end{document}